\documentclass{ws-procs975x65}
\usepackage{wrapft}

\begin{document}

\newcommand{\beq}{\begin{equation}}
\newcommand{\eeq}{\end{equation}}
\newcommand{\beqs}{\begin{eqnarray}}
\newcommand{\eeqs}{\end{eqnarray}}
\newcommand{\zm}{z_{m}}
\newcommand{\ep}{\epsilon}
\newcommand{\tr}{\textrm{Tr}\,}
\newcommand{\condense}{\langle \bar{T}T \rangle}
\newcommand{\Tr}{\hbox{Tr}}
\newcommand{\Ln}{\hbox{Ln}}
\newcommand{\FT}[1]{{\cal FT}#1}
\newcommand{\hyper}[4]{ \hbox{F}({#1},{#2},{#3};{#4}) }  
\newcommand{\half}{{1\over2}}
\newcommand{\VEV}[1]{\langle #1 \rangle}
\newcommand{\bra}[1]{\langle #1 |}
\newcommand{\ket}[1]{| #1 \rangle}
\newcommand{\xSB}{\mbox{$\chi SB$}}
\newcommand{\DxSB}{\mbox{$D\chi SB$}}
\newcommand{\Lag}{\mbox{${\cal L}$}}
\newcommand{\DISP}{\displaystyle}
\newcommand{\intp}[1]{\int {d^4 #1 \over (2\pi)^4}}
\newcommand{\integ}[2]{\int_{#1}^{#2}\!\!}
\newcommand{\integg}[2]{\int_{#1}^{#2}}
\newcommand{\TeV}{\hbox{TeV}}
\newcommand{\GeV}{\hbox{GeV}}
\newcommand{\MeV}{\hbox{MeV}}
\newcommand{\xbar}[1]{#1 \hspace{-5.5pt}/}
\newcommand{\unit}{{\bf 1}}
\newcommand{\SxSB}{\mbox{$S\chi SB$}}
\newcommand{\NJLd}{\mbox{$\hbox{NJL}_{D<4}$}}
\newcommand{\EQ}[1]{(\ref{#1})}
\newcommand{\GB}{ \langle \alpha G_{\mu \nu}^2 \rangle}
\newcommand{\siml}{\hspace{0.3em}\raisebox{0.4ex}{$<$}
\hspace{-0.75em}\raisebox{-.7ex}{$\sim$}\hspace{0.3em}}
\newcommand{\simg}{\hspace{0.3em}\raisebox{0.4ex}{$>$}
\hspace{-0.75em}\raisebox{-.7ex}{$\sim$}\hspace{0.3em}}

\renewcommand{\thefootnote}{\#\arabic{footnote}}
\def\fsl#1{\setbox0=\hbox{$#1$}           
   \dimen0=\wd0                                 
   \setbox1=\hbox{/} \dimen1=\wd1               
   \ifdim\dimen0>\dimen1                        
      \rlap{\hbox to \dimen0{\hfil/\hfil}}      
      #1                                        
   \else                                        
      \rlap{\hbox to \dimen1{\hfil$#1$\hfil}}   
      /                                         
   \fi}       
\title{CONFORMAL HIGGS, OR TECHNI-DILATON
\\ \quad \quad \quad- COMPOSITE HIGGS NEAR CONFORMALITY
}

\author{KOICHI YAMAWAKI}
\address{Depatment of Physics, Nagoya University\\
Nagoya, Japan\footnote{Present Address: Kobayashi-Maskawa Institute for the Origin of Particles and the Universe (KMI), Nagoya University. E-mail: yamawaki@kmi.nagoya-u.ac.jp}
}
\begin{abstract}
In contrast to the folklore that Technicolor (TC) is a ``Higgsless theory'',
we shall discuss existence of  a composite Higgs boson,  Techni-Dilaton (TD),  a pseudo-Nambu-Goldstone boson of the scale invariance 
in the Scale-invariant/Walking/Conformal TC (SWC TC) which generates  a large anomalous dimension $\gamma_m \simeq 1$ in a wide region from the dynamical mass  $m$ 
 $= {\cal O}$  (TeV) of the techni-fermion
all the way up to the intrinsic scale $\Lambda_{\rm TC}$ of the SWC TC  (analogue of $\Lambda_{\rm QCD}$), 
where $\Lambda_{\rm TC} $ is taken typically as the scale of the Extended TC scale $\Lambda_{\rm ETC}$: $\Lambda_{\rm TC} \simeq \Lambda_{\rm ETC}\sim10^3$ TeV
$ (\gg m)$.  
All the techni-hadrons have mass on the same order ${\cal O} (m)$, which in SWC TC is extremely smaller than the intrinsic scale $\Lambda_{\rm TC} \simeq \Lambda_{\rm ETC} $, in sharp contrast to QCD where both are of the same order.  The mass of TD arises from the {\it non-perturbative scale anomaly} associated with the techni-fermion mass generation and is  typically 500-600 GeV,  even {\it smaller than other techni-hadrons} of the same order of  ${\cal O} (m)$,
in another contrast to QCD which is believed to have no scalar $\bar q q$ bound state lighter than other hadrons.
We discuss the TD mass in various methods, Gauged NJL model via ladder Schwinger-Dyson (SD) equation, straightforward  
calculations in the ladder SD/ Bethe-Salpeter equation,  and the holographic approach including techni-gluon 
condensate. The TD may be discovered in LHC.
\end{abstract}


\bodymatter

\section{Introduction}\label{aba:sec1}
Toshihide Maskawa is famous for 2008 Nobel prize-winning paper with Makoto Kobayashi on CP violation
but  did also fundamental contributions particularly  to the SCGT, the topics of  this workshop: Back in 1974 he found
with Hideo Nakajima\cite{Maskawa:1974vs} 
that  spontaneous chiral symmetry breaking  (S$\chi$SB) solution does exists for and only for the strong gauge coupling, with the critical coupling of order 1,
based on 
the ladder Schwinger-Dyson (SD) equation with non-running (scale-invariant) coupling, namely the walking gauge dynamics what is called  today.
This turned out to be the origin of  SCGT activities toward understanding the Origin of Mass.
The present workshop SCGT 09 was held in honor of his 70th birthday  on  February 7, 2010 and the 35th anniversary of his crucial contributions to SCGT.
 I will later explain impact of Maskawa-Nakajima solution on the conformal gauge dynamics. 

The Origin of Mass is the most urgent issue of the particle physics today and is 
to be resolved at the LHC experiments. 
In the standard model (SM),  all masses are attributed to a single parameter of the vacuum expectation value (VEV),  $\langle H \rangle$ of 
the hypothetical elementary particle, the Higgs boson. The VEV  simply picks up the mass scale of the input parameter $M_0$ 
which is tuned to be tachyonic ($M_0^2<0$) in such a way as to tune $\langle H \rangle \simeq 246 \,{\rm GeV}$  (``naturalness problem''). 
As such SM does not explain the Origin of Mass.

 Technicolor (TC)~\cite{TC} is an attractive idea to account for the Origin of Mass 
 without introducing ad hoc Higgs boson and tachyonic mass parameter:  The mass arises {\it dynamically} from the condensate of the techni-fermion and the anti techni-fermion pair $\langle \bar T T \rangle$ which is triggered by the attractive gauge forces between the
 pair analogously to the quark-antiquark condensate $\langle \bar q q \rangle $
  in QCD.    
 For the TC with $SU(N_{\rm TC})$ gauge symmetry and $N_f$ flavors ($N_f/2$ weak doublets) of  techni-fermions,  the techni-pion decay constant $F_\pi=\langle H \rangle/\sqrt{N_f/2}$ 
 corresponds to the pion decay constant $f_\pi \simeq 93\, {\rm MeV}$ in QCD, and hence the TC may be a  scale-up of QCD by the factor $F_\pi/f_\pi \simeq 2650/\sqrt{N_f/2}$.
 Then  the mass scale of the condensate $\Lambda_\chi=(-\langle \bar T T \rangle/N_{\rm TC})^{1/3}$ as the Origin of Mass may be estimated as 
 \beq
 \Lambda_\chi 
 \simeq \left(\frac{- \langle \bar q q \rangle}{N_{\rm c}}\right)^{1/3} 
 \cdot  \frac{F_\pi/\sqrt{N_{\rm TC}}}{ f_\pi/\sqrt{N_{\rm c}}}   \simeq 450 \, {\rm GeV} \cdot\left(\frac{N_{\rm c}/N_{\rm TC}}{N_f/2}\right)^{1/2}\, ,
\eeq
where we have used a typical value $(- \langle \bar q q \rangle)^{1/3}\simeq 250 \, {\rm MeV}\, (N_c=3)$.

 The dynamically generated mass scale of the condensate $\Lambda_\chi$, or the dynamical mass of the techni-fermion,  $m \,(\sim \Lambda_\chi\sim F_\pi)$,  in fact 
 picks up the intrinsic mass scale  $\Lambda_{\rm TC}$ of the theory 
 (analogue of $\Lambda_{\rm QCD}$ in QCD) already generated by the scale anomaly through quantum effects  (``dimensional transmutation'')  in the gauge theory which is scale-invariant at classical level (for massless flavors): 
 \beq
 \Lambda_{\rm TC} = \mu \cdot \exp \left( - \int^{\alpha(\mu)} \frac{d\alpha}{\beta(\alpha)} \right)=\Lambda_0 \cdot \exp \left( - \int^{\alpha(\Lambda_0)} \frac{d\alpha}{\beta(\alpha)} \right) \,,
\label{intrinsic}
\eeq 
where the running of the coupling constant $\alpha(\mu)$, with non-vanishing beta function $ \beta(\alpha) \equiv \mu \frac{d \alpha(\mu)}{d \mu} \ne 0$,
 is a manifestation of the scale anomaly and $\Lambda_0$ is a fundamental scale like 
 Planck scale.  Note that $\Lambda_{\rm TC}$ is independent of the renormalization point $\mu$, $\frac{d \Lambda_{\rm TC}}{d \mu}=0$, 
 and 
 can largely be separated from $\Lambda_0$ through logarithmic running  (``naturalness'').
Thus the Origin of Mass is eventually the quantum effect in this picture: In the simple scale-up of QCD we would have
\beq 
{\rm Naturalness}\,\,({\rm QCD \,\, scale\, up}):\quad \quad \quad  m \sim \Lambda_\chi \sim \Lambda_{\rm TC} \ll \Lambda_0\,.
\label{scaleup}
\eeq

The original version of TC, 
just a simple scale-up 
of QCD, however,  is plagued by the notorious problems: Excessive flavor-changing neutral currents (FCNCs), and
excessive oblique corrections of 
${\cal O}(1)$ to the Peskin-Takeuchi $S$ parameter~\cite{PeskinTakeuchi} compared with the typical
experimental bound about 0.1.  

 The FCNC problem was resolved  long time ago 
 by the TC based on the near conformal gauge dynamics with  $\gamma_m \simeq 1$~\cite{Yamawaki:1985zg, Akiba},
initially  dubbed ``scale-invariant TC"  and then ``walking TC'',  
with almost {\it non-running (conformal) gauge coupling},  based  on the pioneering work by Maskawa and Nakajima~\cite{Maskawa:1974vs} who discovered 
{\it non-zero critical coupling}, $\alpha_{\rm cr} (\ne 0)$, for the S$\chi$SB  to occur. We may call it  ``Scale-invarinat/Walking/Conformal TC'' (SWC TC)
(For reviews see
Ref. \cite{Hill:2002ap}).  

In addition to solving the FCNC problem, the theory made a definite prediction of ``Techni-dilaton (TD)''~\cite{Yamawaki:1985zg}, a pseudo Nambu-Goldstone (NG) boson of the spontaneous breaking of the (approximate) scale invariance of the theory.    This will be the main topics of this talk in the light of modern version of SWC TC.

\begin{figure}[h]
\begin{center}
 \includegraphics[width=5cm]{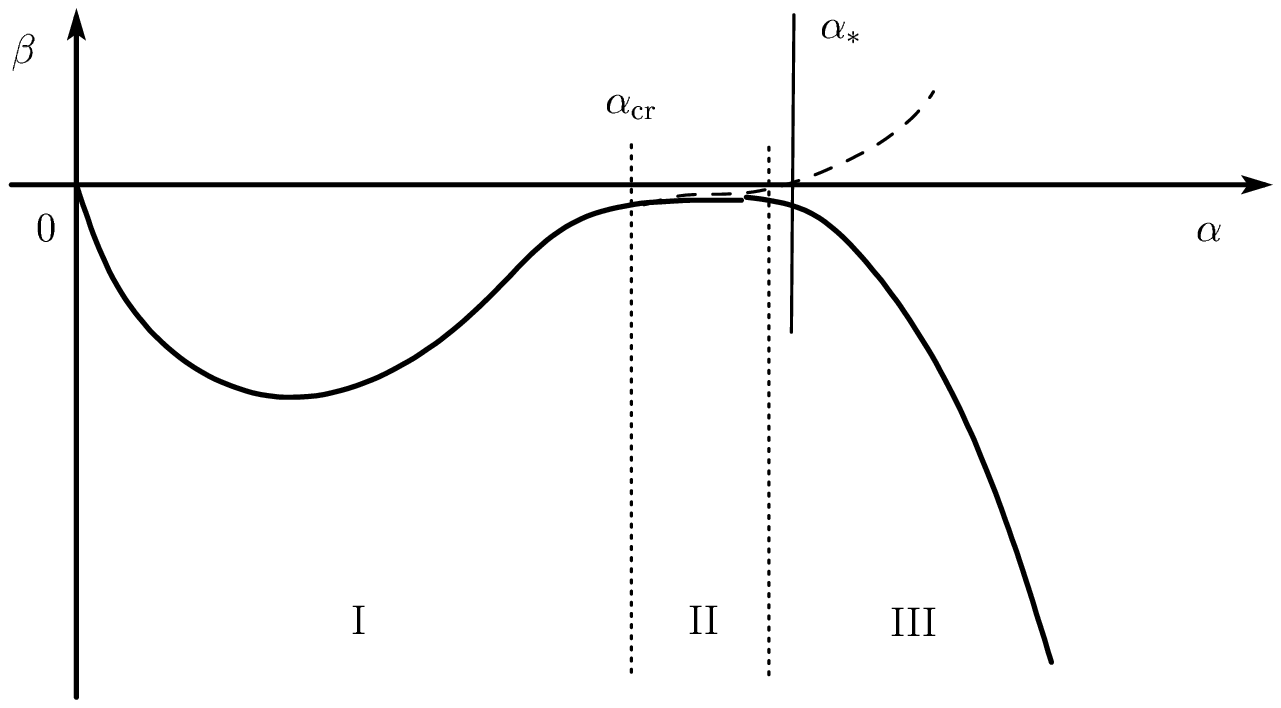}
 \includegraphics[width=6cm]{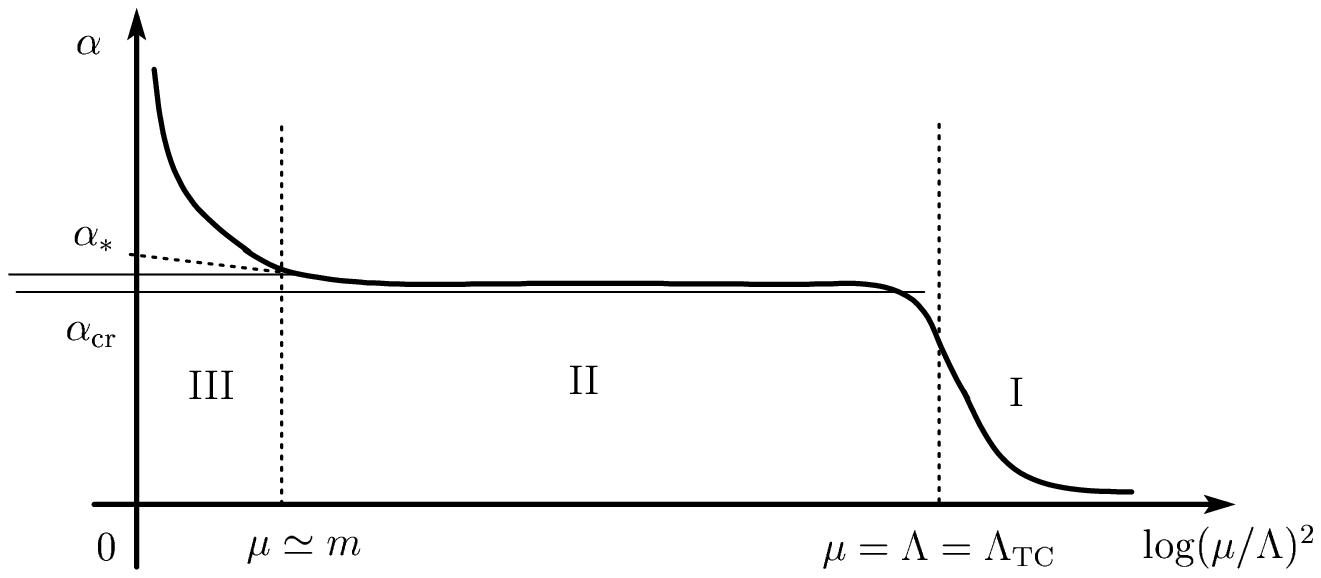} 
\caption{\label{fig:beta}
The beta function and $\alpha(\mu)$ for SWC-TC.}
\end{center} 
\end{figure}

The modern version~\cite{Lane:1991qh,Appelquist:1996dq, Miransky:1996pd}  of SWC TC is based on the Caswell-Banks-Zaks (CBZ) infrared (IR)  fixed point ~\cite{Caswell:1974gg} ,
$\alpha_* =\alpha_*(N_f,N_{\rm TC})$,   which appears at two-loop beta function for the number of massless 
flavors $N_f (<11N_{\rm TC}/2)$ larger than a certain number $N_f^* (\gg N_{\rm TC})$.  
See Fig. \ref{fig:beta} and later discussions.
Due to the IR fixed point the coupling is almost non-running  (``walking'') all the way up to the intrinsic scale 
 $\Lambda_{\rm TC}$ which  is generated by the  the scale anomaly associated with  the (two-loop) running of the coupling
analogously to QCD scale-up in Eq. (\ref{intrinsic}).  For $\mu>\Lambda_{\rm TC}$  (Region I of  Fig.~\ref{fig:beta})  the coupling no
longer walks and runs similarly to that of QCD. 
When we set $\alpha_*$ slightly larger than $\alpha_{\rm cr}$, we have a condensate or the dynamical mass of the techni-fermion $m \, (\sim \Lambda_\chi)$, 
much smaller than
the intrinsic scale of the theory $m \ll \Lambda_{\rm TC} $. 
The  CBZ-IR fixed point  $\alpha_*$ actually disappears (then becoming would-be IR fixed point) 
at the scale $\mu \siml m$ where the techni-fermions have acquired 
the mass $m$ and get decoupled from the beta function for $\mu<m$ (Region III in Fig.~\ref{fig:beta}). Nevertheless,  the coupling is 
still walking due to the remnant of the CBZ-IR fixed point  conformality 
in a wide region $m < \mu < \Lambda_{\rm TC} $ (Region II in Fig.~\ref{fig:beta}).  Thus the {\it symmetry responsible for the natural hierarchy
$m\sim\Lambda_\chi \ll \Lambda_{\rm TC}$  is the
(approximate) conformal symmetry}, while the naturalness for the hierarchy $\Lambda_{\rm TC} \ll \Lambda_0$ is the same as that of QCD
scale-up 
in Eq.(\ref{scaleup}):  
\beq
 {\rm Naturalness}\,\, ({\rm SWC\, TC}) :\quad \quad\quad  m\sim \Lambda_\chi \ll \Lambda_{\rm TC}\quad(\ll \Lambda_0)\, .
\eeq
The theory acts like the SWC-TC~\cite{Yamawaki:1985zg, Akiba} : 
It develops a large anomalous dimension $\gamma_m \simeq 1$ for the almost non-running coupling in the Region II
~\cite{Appelquist:1996dq, Miransky:1996pd}. Here $\Lambda_{\rm TC}$ plays a role of 
cutoff $\Lambda$  identified with the ETC scale: $\Lambda_{\rm TC} =\Lambda=\Lambda_{\rm ETC}$. 

Moreover, there also exists a possibility~\cite{Sundrum:1991rf,Harada:2005ru} that the $S$ parameter can be reduced in the case of SWC-TC.  

In this talk 
I will argue~\cite{Haba:2010iv} \footnote{Preliminary discussions on the revival of the techni-dilaton~\cite{Yamawaki:1985zg}  were given in several talks~\cite{Yamawaki:2009vb} .
}
 that in contrast to the simple QCD scale-up which is widely believed to have no composite  Higgs particle (``higgsless''),  
a salient feature of SWC TC  is the {\it conformality which manifests itself  by the appearance of  a composite Higgs boson (``conformal Higgs'') as the Techni-dilaton (TD)}~\cite{Yamawaki:1985zg}  with
mass relatively lighter than other techni-hadrons: $M_{\rm TD} 
< M_\rho, M_{a_1}\cdots =
 {\cal O}(\Lambda_\chi) \ll \Lambda_{\rm TC}=\Lambda_{\rm ETC}$, where  $M_\rho, M_{a_1}\cdots$ denote the mass of techni-$\rho$, techni-$a_1$, etc. This
is contrasted to the QCD dynamics where there are no scalar bound states lighter than others. 
Note that there is no idealized  limit where the TD  becomes exactly massless to be a true NG boson,  in sharp contrast to the chiral symmetry breaking. 
Scale symmetry is always broken explicitly as well as spontaneously~\footnote{ The straightforward calculations near the conformal edge 
indicated~\cite{Harada:2003dc} that  there is
no isolated massless spectrum: $M_{\rm TD}/F_\pi, M_{\rm TD}/M_\rho, \cdots \to {\rm const.} \ne 0$ even in the limit of  
 $\alpha_*  \to \alpha_{\rm cr}$ ($N_f \to N_f^{\rm crit}$)
where $F_\pi/\Lambda_{\rm TC},
M_{\rm TD}/\Lambda_{\rm TC} ,  M_\rho/\Lambda_{\rm TC} , \cdots  \to 0$.
In the case of holographic TD~\cite{Haba:2010iv},  this fact is realized in a different manner:  Although there apparently  exists  an isolated massless spectrum, 
 $M_{\rm TD}/F_\pi \to  0$ while $M_\rho/F_\pi, M_{a_1}/F_\pi  \to {\rm const.} \ne 0$, 
the decay constant of the TD diverges $F_{\rm TD}/F_\pi \to \infty$ in that limit and hence it  gets decoupled. See later discussions.
} .
 
For the phenomenological purpose,  I will argue through  several different calculations~\cite{Shuto:1989te,Harada:2003dc,Haba:2010iv}
 that  the techni-dilaton mass in the typical SWC TC models 
will be in the range  (see the footnote below Eq.(\ref{prediction}), however): 
\beq
m_{\rm TD} 
= 500-600  \,{\rm GeV}, 
\eeq
which is definitely larger than the SM Higgs bound but still  within the discovery region of the LHC experiments.

\section{Scale-invariant/Walking/Conformal Technicolor}\label{aba:sec2}

Let us briefly review the SWC TC.

The FCNC problem is related with the mass generation of quarks/leptons mass. In order to communicate  the techni-fermion condensate to the
quarks/leptons masses $m_{q/l}$, we would need  interactions between the quarks/leptons and the techni-fermions which are typically introduced through Extended TC (ETC)~\cite{Dimopoulos:1979es}~\footnote{
The same can be done in a composite model where quarks/leptons and techni-fermions are composites on the same footing.\cite{Yamawaki:1982tg}
}   with much higher scale $\Lambda_{\rm ETC} (\gg \Lambda_\chi)$: $m_{q/l} \sim \frac{-1}{\Lambda_{\rm ETC}^2}\, \langle \bar T T\rangle_{\Lambda_{\rm ETC}}$,
 where $\langle \bar T T \rangle_{\Lambda_{\rm ETC}}$ is the condensate measured at the scale of $\Lambda_{\rm ETC}$. (We here do not refer to the origin of  the
 mass scale $\Lambda_{\rm ETC}$ which should also be of dynamical origin such as the tumbling. ) 
Since the newly introduced ETC interactions characterized by the same scale $\Lambda_{\rm ETC}$ should  induce extra FCNC's,  we should impose a constraint $\Lambda_{\rm ETC} > 10^{6} {\rm GeV} $ 
 in order to avoid the  excessive FCNC's (typically involving $s$ quark). 
If we assume
a simple QCD scale up,  $\langle \bar T T \rangle_{\Lambda_{\rm ETC}} 
\simeq \langle \bar T T \rangle_{\Lambda_\chi} =  - N_{\rm TC} \cdot \Lambda_\chi^3 $,
 we would have  
\beq
 m_{q/l} \sim  \frac{\Lambda_\chi^3}{\Lambda_{\rm ETC}^2} \cdot N_{\rm TC} <   
0. 1\, {\rm MeV} \, \cdot N_{\rm TC} \left(\frac{N_{\rm c}/N_{\rm TC}}{N_d}\right)^{3/2}\, ,
\label{qlmass}
\eeq
which implies that  the typical mass ($s$-quark mass) would be roughly $10^{-3}$ smaller than the reality.  We would desperately need $10^3$ times enhancement.

\begin{wrapfigure}{r}{6.6cm}
\vspace{-0.5cm}
\includegraphics[width=6cm]{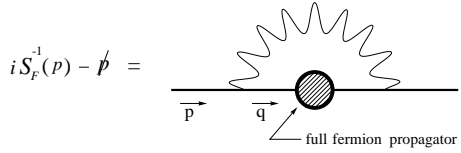}
\caption{Graphical expression of the SD equation 
        in the ladder approximation.}
\label{fig:SDeq}
\end{wrapfigure}

 This was actually realized dynamically  
 by the TC based on the near conformal gauge dynamics~\cite{Yamawaki:1985zg, Akiba}, based  on  the  Maskawa-Nakajima solution~\cite{Maskawa:1974vs} of
 the (scale-invariant) ladder Schwinger-Dyson (SD) equation for fermion full propagator $S_F(p)$ parameterized as
 $i S_{F}^{-1}(p) = A(p^2) \fsl{p} - B(p^2)$  with {\it non-running}  (conformal, an ideal limit
of the ``walking'')  gauge coupling, $\alpha(Q) \equiv  \alpha = {\rm constant}$, with $Q^2\equiv -p^2>0$. (See Fig. \ref{fig:SDeq}) 

 Maskawa and Nakajima 
 discovered that the S$\chi$SB can only take place for strong coupling $\alpha>\alpha_{\rm cr} 
={\cal O} (1) $,  {\it non-zero critical coupling}.\footnote{
Earlier works\cite{JBW} in the ladder SD equation with non-running coupling all confused explicit
breaking solution with the SSB solution and thus implied $\alpha_{\rm cr}=0$}  The critical value reads~\cite{Fukuda:1976zb}:
 $C_2(F) \alpha_{\rm cr}=\pi/3$,  or
 \beq
 \alpha_{\rm cr}=(\pi/3)\cdot 2N_{\rm TC}/(N_{\rm TC}^2-1) 
 \label{critical}
 \eeq
in the $SU(N_{\rm TC}) $ gauge theory,  where $C_2(F)$ is the quadratic Casimir of
the techni-fermion representation of the TC. 
The asymptotic  form of the Maskawa-Nakajima S$\chi$SB solution of the fermion mass function
 $\Sigma(Q) =B(p^2)/A(p^2)$ in Landau gauge ($A(p^2)\equiv 1$) reads~\cite{Maskawa:1974vs,Fukuda:1976zb}, 
 \begin{eqnarray}
\Sigma (Q) \sim 1/Q \quad (Q 
 \gg \Lambda_\chi) \,. 
\label{asymp}
\end{eqnarray}

We then proposed a ``Scale-invariant TC'' ~\cite{Yamawaki:1985zg} , based on  the observation that Eq.(\ref{asymp}) implies a special value of the anomalous dimension
\begin{eqnarray}
\gamma_m = - \Lambda \frac{\partial \ln Z_m}{\partial \Lambda} = 1,  
\label{gammam}
\end{eqnarray}
to be compared  with the operator product expansion (OPE), $\Sigma (Q) \sim 1/Q^2\cdot (Q/\Lambda_\chi)^{\gamma_m}$.
Accordingly, we had an enhanced condensate   $\langle \bar{T} T \rangle_{\Lambda_{\rm ETC}} =   Z_m^{-1}\cdot \langle \bar{T} T \rangle_{\Lambda_\chi}
\simeq -N_{\rm TC} (\Lambda_{\rm ETC} \Lambda_\chi^2)$, with the (inverse) mass renormalization constant  being $Z_m^{-1} = (\Lambda_{\rm ETC}/\Lambda_\chi)^{\gamma_m}\simeq
\Lambda_{\rm ETC}/\Lambda_\chi \simeq 10^3$, which in fact  yields the desired enhancement.  
We actually obtained a different  formula than Eq.(\ref{qlmass}):\cite{Yamawaki:1985zg}  
\beq
 m_{q/l} \sim \frac{\Lambda_\chi^2}{\Lambda_{\rm ETC}}\cdot N_{\rm TC}  \,. 
\eeq
The model~\cite{Yamawaki:1985zg}  was formulated  in terms of the  Renormalization Group Equation (RGE) a la Miransky ~\cite{Miransky:1984ef}  
for the 
Maskawa-Nakajima solution $\Sigma(m)=m$ which takes the form~\cite{Fukuda:1976zb, Miransky:1984ef}  
 \beq
\Lambda_\chi \sim m \sim 4 \Lambda \, \exp \left(- \pi/\sqrt{\alpha/\alpha_{\rm cr} -1}
\right) \, ,
\label{Miransky}
\eeq  
 where $\Lambda$ is the cutoff of the SD equation. This has an {\it essential singularity} often called  ``Miransky scaling''  and 
implies  the {\it non-perturbative beta function} having a {\it multiple zero}~\footnote{
Simple zero of the beta function, $\beta(\alpha) \sim (\alpha-\alpha_{\rm cr})^1$,  never reproduces the essential singularity scaling, as is evident from Eq.(\ref{intrinsic}). 
}
:
\beq
\beta(\alpha)_{_{\rm NP}}= \Lambda \frac{\partial \alpha(\Lambda)}{\partial \Lambda}= -\frac{2}{3C_2(F)}\left( \frac{\alpha}{\alpha_{\rm cr}}-1 \right)^{3/2}\, ,
\label{Miranskybeta}
\eeq
with  
 the critical coupling $\alpha_{\rm cr}$ identified with a nontrivial ultraviolet (UV)  stable fixed point 
 $\alpha=\alpha(\Lambda)  \rightarrow \alpha_{\rm cr}$
as $\Lambda/m \rightarrow \infty$. 

Subsequently,  similar enhancement effects of the condensate were also studied~\cite{Akiba} within the same framework of the ladder SD equation, without use of the RGE concepts of anomalous dimension and fixed point, rather emphasizing 
the asymptotic freedom of the TC theories with slowly-running (walking) coupling which was implemented into the ladder SD equation (``improved ladder SD equation'').  

Today the Scale-invariant/Walking/Conformal TC (SWC TC) is simply characterized by near conformal property 
with  $\gamma_m \simeq 1$ (For a review see
Ref. \cite{Hill:2002ap}).  
Such a theory should have an almost non-running  and strong gauge coupling  (larger than a certain non-zero critical coupling for S$\chi$SB)  to be realized
either at UV fixed point  
or  IR fixed point, or  both (``fusion''  of  the IR and UV fixed points),  as was characterized by 
``{\it Conformal Phase Transition (CPT)}''~\cite{Miransky:1996pd}. 

The essential feature of the above is precisely what happens in the 
modern version ~\cite{Lane:1991qh,Appelquist:1996dq,Miransky:1996pd}  of the SWC TC
 based on the CBZ IR fixed point~\cite{Caswell:1974gg}  of  the large $N_f$ QCD,   the QCD-like theory with many  flavors $N_f  \, (\gg N_{\rm TC})$ of  massless 
 techni-fermions, \footnote{
For SWC TC based on higher representation/other gauge groups see, e.g.,  Ref. \cite{Sannino:2004qp} 
}  see Fig. \ref{fig:beta}.
The  two-loop beta function is given by
 $\beta(\alpha) =\mu \frac{d}{d \mu} \alpha(\mu) 
  = -b \alpha^2(\mu) - c \alpha^3(\mu) ,
$
where
$  b = \left( 11 N_{\rm TC} - 2 N_f \right)/(6 \pi)$, 
 $ c =\left[ 34 N_{\rm TC}^2 - 10 N_f  N_{\rm TC} 
      - 3 N_f  (N_{\rm TC}^2 - 1)/N_{\rm TC} \right] /(24 \pi^2)$ .
When $b>0$ and $c<0$, i.e.,  $  N_f^* < N_f < \frac{11}{2} N_{\rm TC} $ 
($N_f^\ast \simeq 8.05$ for $N_{\rm TC} = 3$), there exists an IR fixed point (CBZ  IR fixed point)
  at $\alpha=\alpha_*$, $\beta(\alpha_*)=0$, where
\begin{equation}
  \alpha_\ast =\alpha_*(N_{\rm TC}, N_f)  = - b/c .
\label{eq:alpha_IR}
\end{equation}
Note that $\alpha_* =\alpha_*(N_f,N_{\rm TC} ) \rightarrow 0$ as $N_f \rightarrow  11 N_{\rm TC}/2$  $ (b \to 0)$ and hence there exists a certain range $N_f^{\rm cr} <N_f < 11 N_{\rm TC}/2$  (``Conformal Window'')  satisfying $\alpha_* < \alpha_{\rm cr}$, where the gauge coupling $\alpha(\mu) \, (< \alpha_*)$ gets so weak
that attractive forces are no longer strong enough to trigger the S$\chi$SB as was demonstrated by Maskawa-Nakajima~\cite{Maskawa:1974vs}.
$N_f^{\rm cr}$ such that  $\alpha_*(N_{\rm TC}, N_f^{\rm cr}) = \alpha_{\rm cr}$ may be evaluated by using  the value of
$\alpha_{\rm cr}$ from  the ladder SD equation Eq.(\ref{critical}): ~\cite{Appelquist:1996dq}  
$N_f^{\rm cr} \simeq 4 N_{\rm TC}$ ($ =12$ for $N_{\rm TC}=3$)
\footnote{
The value should not be taken seriously, since $\alpha_\ast=\alpha_{\rm cr}$ is of  $\cal {O} $(1) and
the perturbative estimate of  $\alpha_*$  is not so reliable there, although the chiral symmetry restoration in large $N_f$ QCD has been supported by many other arguments, most notably the lattice QCD simulations~\cite{lattice,Appelquist:2009ty}, which however
 suggest diverse results as to $N_f^{\rm cr}$; See e.g., \cite{SCGT09} for recent results.
}.

Here we are interested in the S$\chi$SB phase slightly off the conformal window, 
$0< \alpha_\ast - \alpha_{\rm cr}\ll 1$ ($N_f \simeq N_f^{\rm cr}$).  
We may use the same equation as the ladder SD  equation  with $\alpha(\mu) \simeq {\rm const.} =\alpha_*$, yielding 
 the
same form as 
Eq.(\ref{Miransky})
:~\cite{Appelquist:1996dq}
\begin{equation}
m 
\sim 4 \Lambda_{\rm TC} \, \exp \left(- \pi/\sqrt{\alpha_*/\alpha_{\rm cr} -1}
\right) \ll  \Lambda_{\rm TC}   \quad ( \alpha_\ast \simeq \alpha_{\rm cr})\, ,
\label{Appel}
\end{equation}
where the cutoff $\Lambda$ was identified with $\Lambda_{\rm TC} (=\Lambda_{\rm ETC}$).  
We also have the same result as Eqs. (\ref{asymp}),(\ref{gammam}): 
\begin{eqnarray}
\Sigma(Q) \sim 1/Q \,,\quad \quad
\gamma_m \simeq  1\,. 
\end{eqnarray}
Hence it acts like SWC TC. Incidentally, Eq.(\ref{Appel}) implies  a {\it multiple zero} at $\alpha_*=\alpha_{\rm cr}$
 in a non-perturbative beta function for $\alpha_*=\alpha_*(\Lambda)$
similar to Eq.(\ref{Miranskybeta}), which would suggest  ``running'' of the IR fixed point  $\alpha_*$ with its
UV fixed point $\alpha_*=\alpha_{\rm cr}$ in the  limit $\Lambda_{\rm TC}/m \rightarrow \infty$.

The actual running of the coupling largely based on two-loop perturbation is already depicted in Fig.~\ref{fig:beta}.
The critical coupling $\alpha_{\rm cr}$ can be regarded as the UV fixed point viewed from the IR part of the Region II ($m< \mu<\mu_{\rm cr}$, with
$\mu_{\rm cr}$ such that $ \alpha(\mu_{\rm cr})=\alpha_{\rm cr}$), while it is regarded as  the IR fixed point
from the UV part of the Region II ($\Lambda_{TC}> \mu >\mu_{\rm cr}$), with the Region II regarded as the {\it fusion of the IR and UV fixed points} in the idealized limit of non-running (perturbative) coupling in Region II
(or $\Lambda_{\rm TC}/m \to \infty$). Although the perturbative (two-loop) beta function has a {\it simple zero}, which {\it never corresponds to the essential singularity} scaling as we noted before,
the coupling  near $\alpha_{\rm cr}$ should be
sensitive to the non-perturbative effects in such as way that the beta function looks like the {\it multiple zero} as in  Eq. (\ref{Miranskybeta})  from both sides,
corresponding to the {\it essential singularity scaling} as  in Eq. (\ref{Miransky}).  This should be tested by the fully non-perturbative studies like lattice simulations.
A possible phase diagram (Fig 3 of Ref.\cite{Miransky:1996pd}) of  the  large $N_f$ QCD on the lattice  is  also waiting for the test by simulations.

\section{Conformal Phase Transition~\cite{Miransky:1996pd,Yamawaki:2009vb}}
\label{CPT}

Such an {\it essential singularity}  scaling law  like Eq.(\ref{Miransky}),(\ref{Appel}), or equivalently the {\it multiple zero} of the non-perturbative beta function,
characterizes an unusual phase transition, what we called ``{\it Conformal Phase Transition (CPT)}'',
where the Ginzburg-Landau effective theory breaks down~\cite{Miransky:1996pd}: Although it is a second order (continuous) phase transition where the order parameter 
$m$ ($\alpha_*> \alpha_{\rm cr}$) is continuously changed to $m=0$ in the symmetric phase (conformal window, $\alpha_* < \alpha_{\rm cr}$), the spectra do not, i.e., while there exist light composite particles whose mass  vanishes at the critical point when approached from the side of  the SSB phase, 
no isolated light particles do not exist 
in the conformal window, recently dubbed  ``unparticle''~\cite{Georgi:2007ek}. 
 This reflects the feature of the conformal symmetry in the conformal window. 
In fact explicit computations show 
no light (composite) spectra in the conformal window,  in sharp contrast to the S$\chi$SB phase
where light composite spectra do exist with mass of order ${\cal O} (m)$ which vanishes as we approach the conformal
window $N_f \nearrow N_f^{\rm cr} $ \cite{Appelquist:1996dq,Miransky:1996pd, Harada:2003dc}.  

The essence of CPT was illustrated ~\cite{Miransky:1996pd} by a simpler model, 2-dimensional Gross-Neveu Model.
This is the $D\rightarrow 2$ limit of the $D$-dimensional Gross-Neveu model  ($2<D<4$) which has 
 the beta function and the anomalous dimension:~\cite{Kikukawa:1989fw, Kondo:1992sq}
\begin{eqnarray}
\beta(g) =-2g (g-g_*),\quad
\gamma_m = 2 g \, ,
\label{betagammaDNJL}
\end{eqnarray}
where  $g=g_*(\equiv D/2-1)=g_{\rm cr}$ and $g=0$ are respectively the UV and IR fixed points of the 
dimensionless four-fermion coupling, $g$, properly normalized (as  $g_* =1$ for  the $D=4$ NJL model).
There exist light composites $\pi, \sigma$ near the UV fixed point (phase boundary) $g\simeq g_*$ in both sides of symmetric ($0<g<g_*$) and SSB 
($g>g_*$)  phases as in the NJL model. 

Now we consider   $D\rightarrow 2$ ( $g_* \rightarrow 0$) where 
we have a well-known effective potential: 
$V(\sigma , \pi) \sim (1/g-1)  \rho^2 +\rho^2 
 \ln (\rho^2/\Lambda^2) $,
or  $\partial^2 V/\partial \rho^2|_{\rho=0} = -\infty$, where $\rho^2=\pi^2+\sigma^2$. This implies breakdown of the Ginzburg-Landau theory which 
 distinguishes the SSB ($<0$) and  symmetric ($>0$) phases by the signature of the finite $\partial^2 V/\partial \rho^2 $ at the critical point $g=0$. 
Eq. (\ref{betagammaDNJL} ) now reads:
\begin{eqnarray}
\beta(g) =-2 g^2\, ,\, \,\,\gamma_m|_{g=0}= 0  \quad \quad (D= 2)\,,
\end{eqnarray}
namely  a {\it fusion of the
UV and IR fixed points} at $g=0$ as a result of {\it multiple zero} (not a simple zero) at $g=0$. 
Now the symmetric phase is squeezed out to the region  $g<0$ (conformal phase) which corresponds to a {\it repulsive} four-fermion interaction and no composite states exist, while in the SSB phase ($g>0$) there exists a composite state $\sigma$ of  mass $M_\sigma =2 m$ where the dynamical mass
of the fermion is given by 
$m^2 \sim \Lambda^2 \exp(-1/g) \rightarrow 0 \,\, (g\rightarrow + 0)$, which shows an {\it essential singularity} scaling, in accord with the beta function with 
{\it multiple zero},
$\beta (g) = \Lambda \partial g/\partial \Lambda=-2 g^2$. Note the would-be
composite mass in the symmetric phase $|M|^2 \sim  \Lambda^2 \exp(-1/g)\rightarrow \infty \,\, (g\rightarrow -0)$.

Now look at the SWC TC as modeled by the large $N_f$ QCD: When the walking coupling  $\alpha(Q) \simeq \alpha_* $ is close to the critical coupling,  $\alpha_* \simeq \alpha_{\rm cr} $,
we should include the {\it induced}  four-fermion interaction, $(G/2) [\left(\bar \psi  \psi
\right)^2 +\left(\bar \psi  i \gamma_5 \psi\right)^2]$, which becomes relevant operator due to the anomalous dimension  $\gamma_m =1$, and  the system becomes 
``gauged Nambu-Jona-Lasinio'' model \cite{Bardeen:1985sm} whose solution in the full parameter space
was obtained in Ref.\cite{Kondo:1988qd}  

Thus we may regard the  {\it SWC TC as the gauged Nambu-Jona-Lasinio (NJL) model}. It was found \cite{Kondo:1988qd}
that  S$\chi$SB solution exists for the parameter space 
$g> g_{(+)}= (1+ \sqrt{1-\alpha_*/\alpha_{\rm cr}})^2/4\,\, \,\,(\alpha_*<\alpha_{\rm cr}) $ as well as the region $\alpha_*>\alpha_{\rm cr}$, where the dimensionless four-fermion coupling $g\equiv G\Lambda^2 \,(N_{\rm TC}/4\pi^2) $ is normalized as $g=1$ for $\alpha_*=0$ (pure NJL model without gauge interaction).
Based on the solution (including the running coupling case), the RGE flow in $(\alpha,g)$ space was found to be along the  line of  $\alpha =\alpha_*$
($\alpha$ does not run), \footnote{
The beta function in Eq.(\ref{Miranskybeta}) may be regarded as  an artificial one keeping $g\equiv {\rm const.}$
which is not along the renormalized trajectory in the extended parameter space $(\alpha,g)$.  
}
 on which  the four-fermion coupling $g$
runs, with the beta function and anomalous dimension given by ~\cite{Kondo:1991yk,Kondo:1992sq,Aoki:1999dv}
 \begin{eqnarray}
\beta(g) =-2(g-g_{(+)})(g-g_{(-)}), \quad 
\gamma_m = 2g +
\alpha_*/(2\alpha_{\rm cr})
\label{betagammaGNJL}
\end{eqnarray}
where    $g=g_{(\pm)}\equiv 
(1\pm \sqrt{1-\alpha_*/\alpha_{\rm cr}})^2/4$ are regarded as the UV/IR fixed 
points (fixed lines) for $\alpha_*\le \alpha_{\rm cr}$. The above anomalous dimension takes the values:  $\gamma_m= 1+ \sqrt{1-\alpha_*/\alpha_{\rm cr}} $ ~\cite{Miransky:1988gk} at the UV fixed line while $\gamma_m= 1- \sqrt{1-\alpha_*/\alpha_{\rm cr}} $ at the IR fixed line. 
 Light composite spectra only exist near the UV
fixed line (phase boundary) $g\simeq g_{(+)}$ in both SSB ($g>g_{(+)}$) and symmetric ($g>g_{(+)}$) phases as in NJL model. Thus it follows  that as $\alpha_*\rightarrow \alpha_{\rm cr}$ Eq. (\ref{betagammaGNJL})  takes the form
\begin{eqnarray}
\beta(g) = -2 (g-g_*)^2 \,, \,\,\,\gamma_m |_{g=g_*}= 1\, , \quad\quad ( \alpha_* = \alpha_{\rm cr} ) \, ,
\end{eqnarray}
with $g_{(\pm)}\rightarrow 1/4 \equiv g_*$, and hence we again got a {\it multiple zero} and {\it  fusion of UV and IR fixed lines} ~\cite{Kondo:1991yk,Kondo:1992sq,Aoki:1999dv} which  corresponds to
the {\it essential singularity} scaling;~\cite{Kondo:1988qd}  $m^2 \sim \Lambda^2 \exp (-1/(g-g_*)) $.  A similar observation was also made recently.~\cite{Kaplan:2009kr}

In passing, it should be stressed that the {\it anomalous dimension never changes discontinuously across the phase boundary} as is seen  from Eq.(\ref{betagammaDNJL}) and Eq.(\ref{betagammaGNJL}).~\cite{Kikukawa:1989fw, Kondo:1991yk,Kondo:1992sq}

The scale anomaly in this case is given by\cite{Miransky:1996pd}:
\begin{eqnarray}
\VEV{\partial^\mu D_\mu} 
= \VEV{\theta^{\mu}_{\mu}} 
=4 \VEV{ \theta_{0}^{0}}
\nonumber
&=& \frac{\beta (g)}{g} \cdot
\frac{G}{2}\langle\left(\bar \psi  \psi
\right)^2 +\left(\bar \psi  i \gamma_5 \psi\right)^2\rangle \\
&\simeq&- m^4\cdot  (4 N_f N_{\rm TC}/\pi^4) ={\cal O}(\Lambda_\chi^4)\, ,
\label{anomalyNJL}
\end{eqnarray}
where the second line was from  the explicit computation~\cite{Nonoyama:1989dq}
 of the vacuum energy $\langle \theta^0_0\rangle$ in the limit $\Lambda/m \to \infty$ ($g\to g_*$) at 
$\alpha\equiv\alpha_{\rm cr}$ (The result coincides with the one for $\alpha \to \alpha_{\rm cr}$ with $g\equiv 0$, see Eq.(\ref{npanomaly}).~\cite{Miransky:1989qc}.).
Again there is a composite state this time  having mass~\cite{Shuto:1989te}\beq
M_\sigma \to \sqrt{2} m\,,
\label{sigmamass}
\eeq
as $g\rightarrow g_* + 0$, while there are no composites    $|M|^2 \sim \Lambda^2 \exp (-1/(g-g_*))\rightarrow \infty$ for $g\rightarrow g_* - 0$.
 Eq.(\ref{sigmamass})  is compared with $M_\sigma =2 m$ in the pure NJL case with $\alpha\equiv 0$.
This slightly lighter scalar may be identified with the techni-dilaton in the SWC TC. I will come back to this
later.

The absence of the composites in the symmetric phase $g<g_*$ may be understood as in the 2-dimensional Gross-Neveu model for $g<0$,
namely the {\it repulsive} four-fermion interactions:
From the 
analysis of the RG flow, it was
 argued \cite{Kondo:1991yk}  
that the IR fixed line
$g= g_{(-)}$ is  due to the {\it induced} four-fermion interaction by the walking TC dynamics itself, while deviation from that line, $g- g_{(-)}$,  is  due to  
the  {\it additional} four-fermion interactions, repulsive 
($g<g_{(-)}$) and  attractive  ($g>g_{(-)}$), from UV dynamics other than the TC  (i.e., ETC). It is clear that no light composites exist for repulsive four-fermion interaction $g<g_{(-)}$, which becomes $g<g_*$ at $\alpha_*=\alpha_{\rm cr}$.

\section{S Parameter Constraint}

Now we come to the next problem of TC, so-called $S,T,U$ parameters \cite{PeskinTakeuchi} measuring possible new physics in terms of  the deviation of the LEP precision experiments  from the SM. In particular, $S$ parameter 
excludes the TC as a simple scale-up of QCD which yields $S = (N_f/2)  \cdot \hat{S}$ with
$\hat{S}_{\rm QCD} =0.32 \pm 0.04$. For a typical ETC model with one-family TC, $N_f=8$, \cite{TC}
 we would get $S =
{\cal O} (1)$ which is  much larger than the experiments $S < 0.1$. This is the reason why many people  believe that the TC is dead. However, since the simple scale-up of QCD was already ruled out by the FCNC
as was discussed before, the real problem is whether or not the walking/conformal TC which 
solved the FCNC problem is also consistent with the S parameter constraint above.  There have been many arguments \cite{Sundrum:1991rf,Harada:2005ru} that the $S$ parameter value could be reduced in the walking/conformal TC than in the simple scale-up of  QCD. 
Recently such a reduction has also been argued~\cite{Hong:2006si,Haba:2008nz}  in  a version of  the holographic 
  QCD~\cite{Erlich:2005qh}
deformed to the walking/conformal TC  by  tuning a parameter
to simulate the large anomalous dimension $\gamma_m \simeq 1$.

\begin{figure}[h]
   \begin{center}
     \includegraphics[height=2cm]{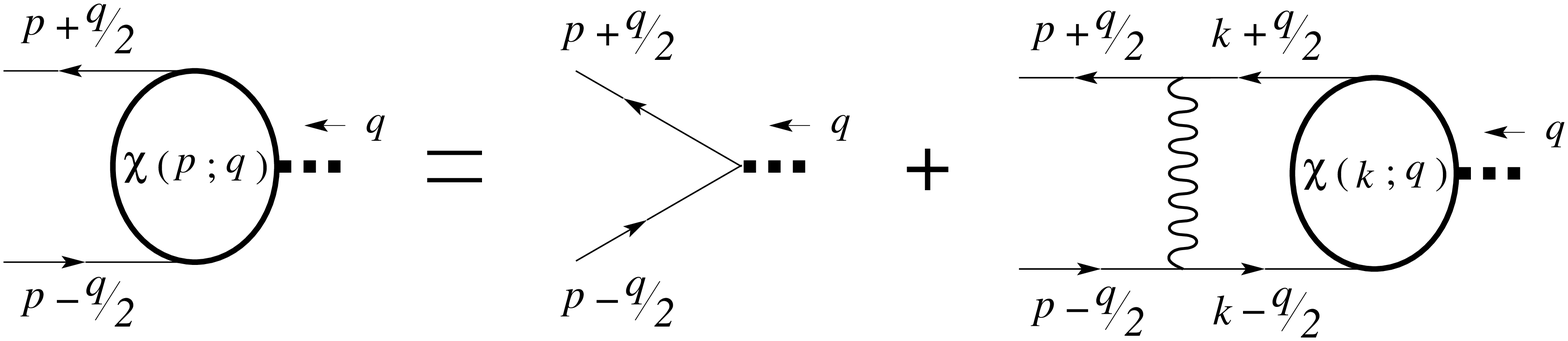}
   \end{center}
\caption{Graphical expression of the BS equation in the 
ladder approximation.}
\label{fig:IBSeq}
 \end{figure}

Here we present the most straightforward computation of the $S$ parameter for the large $N_f$ QCD,
based on the SD equation and (inhomogeneous) 
Bethe-Salpeter (BS) equation in the ladder approximation. \cite{Harada:2005ru}
The $S$ parameter $S=(N_f/2) {\hat S}$ is defined by the slope of the the current correlators $\Pi_{JJ}(Q^2)$ at $Q^2=0$:
\begin{equation}
   {\hat S} =\left.
   - 4 \pi \frac{d}{d Q^2} 
   \left[ \Pi_{VV}(Q^2) - \Pi_{AA}(Q^2)
   \right] \right\vert_{Q^2 = 0} \, ,
 \label{eq:S_parameter}
\end{equation}
where 
  $ \delta^{ab} \left( q_\mu q_\nu/q^2 - g_{\mu \nu} \right) 
 \Pi_{JJ}(q^2) 
= 
\cal{F.T.}  
$
$i \langle 0 \vert T J_\mu^a(x) J_\nu^b(0) 
   \vert 0 \rangle, $ 
$( J_\mu^a(x) = V_\mu^a(x), A_\mu^a(x) ) \, , 
$
with $
   F_\pi^2 = \Pi_{VV}(0) - \Pi_{AA}(0)\, .
$
The current correlators are obtained by closing the fermion legs of the BS amplitudes 
$\chi_\mu^{(J)}
(p;q
) \sim
\cal{F.T.}$
$\langle 0 \vert T\ \psi(r/2)\ 
   \bar\psi
(-r/2)\ J_\mu
(x)\ 
   \vert 0 \rangle$,
   which is determined by the ladder BS equation (Fig.\ref{fig:IBSeq}). 
Solving the BS equation with the
fermion propagator given as the solution of the ladder SD equation, we can evaluate the 
$\Pi_{VV}(Q^2) - \Pi_{AA}(Q^2)$ numerically. From this result we may read its slope  at
$Q^2=0$ to get $\hat{S}$.

The results show  definitely smaller values of   $\hat{S}$  than that in the ordinary QCD and 
moreover there is a tendency of the $\hat{S}$ getting reduced when
approaching the conformal window $\alpha_\ast \searrow \alpha_{\rm cr}$ 
($N_f \nearrow N_f^{\rm cr}$). However,
due to technical limitation of the present computation getting very close to  the conformal window,  the reduction does not seem to be so dramatic as the walking TC being enough to be consistent with the experimental constraints. It is highly desirable to extend the computation further close to the conformal window. 

Another approach to this problem is the deformation of the holographic QCD by the anomalous dimension. 
The reduction of $S$ parameter in the SWC TC has  been argued  in  a version of  the hard-wall type bottom up holographic 
  QCD~\cite{Erlich:2005qh}
deformed to the SWCTC  by  tuning a parameter
to simulate the large anomalous dimension $\gamma_m \simeq 1$. \cite{Hong:2006si} 
We examined \cite{Haba:2008nz} such a possibility paying attention to the renormalization point dependence 
of the condensate. 
We explicitly calculated the $S$ parameter in entire parameter space 
of the holographic SWC TC.  We here take a set of $F_\pi/M_\rho$ and $\gamma_m$. 
We find that $S>0$ and it monotonically decreases to zero in accord with the previous results~\cite{Hong:2006si} .
However, 
 our result turned out  fairly independent of the value of the anomalous dimension $\gamma_m$,  yielding no
particular suppression solely by tuning the anomalous dimension large,  $\hat{S} \sim B (F_\pi/M_\rho)^2 \to 0$ as $F_\pi/ M_\rho \to 0$, 
with $B \simeq 27 (32) $ for $\gamma_m\simeq 1 (0)$,  in sharp contrast to the 
previous claim~\cite{Hong:2006si} .  Although $B$ contains full contributions from the  infinite tower of the vector/axial-vector Kaluza-Klein modes (gauge
bosons of hidden local symmetries)~\cite{Bando:1984ej} of the 5-dimensional gauge bosons,
the resultant value of $B$  turned out  close to $B \simeq 4\pi a \simeq 8\pi $ of the single $\rho$ meson dominance, where $a \simeq 2$  is the parameter of the hidden local symmetry only for the $\rho$ meson.~\cite{Bando:1984ej} This implies that as far as the pure TC dynamics (without ETC dynamics, etc.) is concerned,
an obvious way to dynamically reduce $S$ parameter is to tune $F_\pi/M_\rho$ very small, namely techni-$\rho$ mass very large to several TeV region. (
See, however, footnote below Eq.(\ref{prediction}).)
 
 We would need more dynamical information other than the holographic recipe, since 
the parameter corresponding to $F_\pi/M_\rho$ as well as the scale parameter  is a pure input in all the holographic models, whether bottom up or top down
approach,
in contrast to the underlying gauge theory which has only a single parameter, a scale parameter like $\Lambda_{\rm QCD}$. 

Curiously enough,  
when we calculate  $F_\pi/M_\rho$ from the SD and the homogeneous BS equations~\cite{Harada:2003dc} and 
$S$ from the SD  and the  inhomogeneous BS equation~\cite{Harada:2005ru} both  in the straightforward calculation in the  ladder approximation,
a set of the calculated values of ($F_\pi/M_\rho, S$)  lies on the line of the holographic result~\cite{Haba:2008nz}. 

\section{Techni-dilaton}
 Now we come to the discussions of Techni-dilaton (TD).
Existence of two largely separated scales, 
$\Lambda_\chi\sim m $ and $\Lambda_{\rm TC}$ such that $\Lambda_\chi \ll \Lambda_{\rm TC}$,  
is the most important feature of SWC-TC,  in sharp contrast to 
the ordinary QCD with small number of flavors (in the chiral limit)  
where all the mass parameters like dynamical mass of quarks are of
order of the single scale parameter of the theory $\Lambda_{\rm QCD}$, $m\sim \Lambda_\chi \sim \Lambda_{\rm QCD}$.  See Fig. \ref{fig:beta}.
The intrinsic scale $\Lambda_{\rm TC}$ is related with the scale anomaly corresponding to
the {\it perturbative} running effects
of the coupling, with the ordinary  two-loop beta function 
$\beta (\alpha)$ in the Region I, 
in the same sense as in QCD. 
\beqs
\VEV{\partial^\mu D_\mu} 
= \VEV{\theta^{\mu}_{\mu}} 
= \frac{\beta (\alpha)}{4 \alpha^2} \GB ={\cal O} (\Lambda_{\rm TC}^4), 
\label{panomaly}
\eeqs
which implies that all the techni-glue balls have mass of ${\cal O} (\Lambda_{\rm TC})$.

On the other hand, the scale $\Lambda_\chi$ is related with totally different scale anomaly due to the dynamical generation of $m\, (\sim \Lambda_\chi)$ which does exist 
even in the idealized case with non-running coupling $\alpha(\mu) \equiv \alpha ( >\alpha_{\rm cr})$ such as the Maskawa-Nakajima solution~\cite{Maskawa:1974vs}, as was discussed some time ago~\cite{Miransky:1989qc}. 
Such an idealized case well simulates the dynamics of Region II of Fig. \ref{fig:beta} ~\cite{Appelquist:1996dq, Miransky:1996pd}, with anomalous dimension $\gamma_m \simeq 1$ and $m \ll \Lambda_{\rm TC}$ 
in the numerical calculations~\cite{Harada:2003dc}, with  
the {\it perturbative} coupling constant in Region II being almost constant slightly larger than $\alpha_{\rm cr}$, $\alpha_{\rm cr}< \alpha(\mu)  (< \alpha_*)$,
 for a wide infrared region.  
The coupling $\alpha \equiv \alpha_*$ in the ``idealized Region II" 
actually runs {\it non-perturbatively} according to the {\it essential-singularity} scaling 
(Miransky scaling~\cite{Miransky:1984ef}) of mass generation, Eq.(\ref{Miransky}), 
 with the {\it non-perturbative} beta function $\beta_{\rm NP}(\alpha)$, Eq.(\ref{Miranskybeta}), having a {\it multiple zero} at $\alpha=\alpha_{\rm cr}$.  
Then the {\it non-perturbative} scale anomaly reads~\cite{Miransky:1996pd} ~\footnote{In terms of the gauged NJL model mentioned in Section \ref{CPT} this is the expression of the scale anomaly for $\alpha\to \alpha_{\rm cr}$ with $g={\rm const.}=g_*$, in contrast to Eq.(\ref{anomalyNJL}) for $g\to g_*$ with $\alpha={\rm const.} =\alpha_*=\alpha_{\rm cr}$. Both yield the same vacuum energy and hence the same scale anomaly.
}
\beqs
\VEV{\partial^\mu D_\mu}_{_{\rm NP}} 
= \VEV{\theta^{\mu}_{\mu}}_{_{\rm NP}} 
= \frac{\beta_{_{\rm NP}} (\alpha)}{4 \alpha^2} \GB_{_{\rm NP}}  =- m^4\cdot  \frac{4 N_f N_{\rm TC}}{\pi^4}=- {\cal O}(\Lambda_\chi^4), 
\label{npanomaly}
\eeqs
where $\langle \cdots\rangle_{_{\rm NP}}$ is the quantity with the perturbative contributions subtracted:~\cite{Miransky:1989qc}  $\langle \cdots\rangle_{_{\rm NP}} \equiv  \langle \cdots\rangle - \langle \cdots\rangle_{_{\rm Perturbative}} $.
Eq.(\ref{npanomaly}) coincides with Eq.(\ref{anomalyNJL}) and  $\VEV{\partial^\mu D_\mu}_{_{\rm NP}} /\Lambda_{\rm TC}^4$ vanishes with
$ \VEV{\partial^\mu D_\mu}_{_{\rm NP}}/m^4 \to {\rm const.} \ne 0$,  when we approach the conformal window from the broken phase $\alpha_* \searrow \alpha_{\rm cr}$ ($m/\Lambda_{\rm TC} \to 0$).
All the techni-fermion bound states have mass of order of $m$~\cite{Chivukula:1996kg}, while there are no light bound states
in the symmetric phase (conformal window) $\alpha_*<\alpha_{\rm cr}$, a characteristic feature of the conformal phase transition~\cite{Miransky:1996pd}.
The TD is associated with the latter scale anomaly and should have mass on order of $m (\ll \Lambda_{\rm TC})$.

\subsection{Calculation from Gauged NJL model in the ladder SD equation~\cite{Shuto:1989te}}
More concretely, the mass of TD or scalar bound state in the SWC-TC was estimated in various methods:   
The first method ~\cite{Shuto:1989te}  was based on the 
the  ladder SD equation for the gauged NJL model which well simulates~\cite{Appelquist:1996dq,Miransky:1996pd} the 
conformal phase transition in the large $N_f$ QCD. The result was already given by Eq.(\ref{sigmamass}):
\beqs
M_{\rm TD} \simeq 
\sqrt{2} m \, .
\label{PCDCmass}
\eeqs

\subsection{Straightforward Calculation from Ladder SD and BS equations}

Also a straightforward calculation~\cite{Harada:2003dc} of mass of TD, the scalar bound state 
was made  in the
vicinity of the CBZ-IR fixed point  in the large $N_f$ QCD, based on the coupled use of the ladder SD equation and ({\it homogeneous}) BS equation
lacking the first term in Fig. \ref{fig:IBSeq}:  All the bound states masses are $M = {\cal O}(m)$ and $M/\Lambda_{\rm TC}\to 0$, when approaching 
the conformal window  $\alpha_*\to \alpha_{\rm cr} \,\, (N_f \to N_f^{\rm cr})$ such that $m/\Lambda_{\rm TC} \to 0$.  
Near the conformal window ($N_f \nearrow N_f^{\rm cr}$)  the calculated values are  $M_\rho/F_\pi \simeq 11, M_{a_1}/F_\pi  \simeq 12$ (near degenerate !).
On the other hand, the scalar mass sharply drops near the  
the conformal window,   $M_{\rm TD}/F_\pi \searrow 4$,  or
\beq
M_{\rm TD}\searrow   1.5 m \simeq \sqrt{2} m\,  (< M_\rho, M_{a_1})\, .
\eeq
Note that  in this calculation 
$M_{\rm TD} /F_\pi \to {\rm const.}\ne 0$ and hence  there is no isolated massless scalar bound states even in the limit $N_f \to N_f^{\rm cr}$.
The result  is consistent with Eq.(\ref{PCDCmass}) and is  contrasted to the ordinary QCD where the scalar mass is larger than those of the
 vector mesons (``higgsless'') within the same framework of ladder
 SD/BS equation approach. 
 The result would imply 
 \beq
m_{\rm TD}
\simeq 500\, {\rm GeV}
\eeq
 in the case of the one-family TC model with $F_\pi \simeq 125 \,{\rm GeV}$.

\subsection{Holographic Techni-dilaton~\cite{Haba:2010iv}}
Recently, we have calculated~\cite{Haba:2010iv} mass of TD 
in an extension of  the previous paper~\cite{Haba:2008nz} on the hard-wall-type 
bottom-up holographic SWC-TC by including effects of (techni-) gluon 
condensation parameterized as 
\beqs
\label{B:def}
\Gamma 
\equiv \left(\frac{\left(\frac{1}{\pi}\GB/F_\pi^4\right)}
{\left(\frac{1}{\pi}\GB/f_\pi^4\right)_{\rm QCD}}\right)^{1/4}
\,
\eeqs 
 through the bulk flavor/chiral-singlet scalar field $\Phi_X$, 
in addition to the conventional bulk scalar field $\Phi$ dual to the chiral condensate. 

 The  five-dimensional action is given by 
 \beqs
S_5 &=&\,\int\,d^4 x\,\int_{\ep}^{\zm}\,d\,z~\sqrt{-g}\,
\frac{1}{ g_5^2}e^{c g_5^2 \Phi_X(z)} \Big(
-\frac{1}{4}\tr\left[{L_{MN}L^{MN}}
+{R_{MN}R^{MN}}\right]
\nonumber \\ 
&& 
\hspace{90pt} 
+\tr\left[{D_M\Phi^\dagger D^M\Phi}
-m^2_\Phi \Phi^\dagger \Phi \right]
+ \frac{1}{2} \partial_M 
\Phi_X \partial^M \Phi_X \Big)
\,, 
\label{S5}
\eeqs
where the anti-de-Sitter space (AdS$_5$) with the curvature radius  $L$ of AdS$_5$
is described by the metric $ds^2= g_{MN} dx^M dx^N 
= 
\left(L/z \right)^2\big(\eta_{\mu\nu}dx^\mu dx^\nu-dz^2\big)$ 
with $\eta_{\mu\nu}={\rm diag}[1, -1, -1,-1]$, 
$g={\rm det}[g_{MN}]=-(L/z)^{10}$; 
$g_5$ denotes the gauge coupling in five-dimension 
and $c$ is the dimensionless coupling constant, and $L_M(R_M)=L_M^a(R_M^a) T^a$ with the generators of $SU(N_f)$ are normalized by 
${\rm Tr}[T^a T^b]=\delta^{ab}$; 
$L(R)_{MN} = \partial_M L(R)_N - \partial_N L(R)_M 
 - i [ L(R)_M, L(R)_N ]$. 
The covariant derivative acting on $\Phi$ is defined as 
$D_M\Phi=\partial_M \Phi+iL_M\Phi-i\Phi R_M$.  

The TD, a flavor-singlet scalar bound state of
techni-fermion and anti-techni-fermion, will be identified with the lowest KK mode 
coming from the bulk scalar field $\Phi$, not $\Phi_X$.   
Thanks to the additional explicit bulk scalar field $\Phi_X$, 
we naturally improve the matching with the OPE 
of the underlying theory (QCD and SWC-TC) for current correlators 
so as to reproduce gluonic $1/Q^4$ term, 
which is clearly distinguished from the same $1/Q^4$ terms from 
chiral condensate in the case of SWC-TC with $\gamma_m \simeq 1$. 
Our model with $\gamma_m =0$ and $N_f=3$ 
well reproduces the real-life QCD. 

It is rather straightforward \cite{Erlich:2005qh,Hong:2006si}  to compute masses of the techni-$\rho$ meson ($M_\rho$), 
the techni-$a_1$ meson ($M_{a_1}$),  
while for  that of  the TD, flavor-singlet scalar meson ($M_{\rm TD}$),  we  would need additional IR potential with quartic coupling $\lambda$ to stabilize the S$\chi$SB vacuum~\cite{DaRold:2005vr}.
Such an IR potential  might be regarded as generated by techni-fermion loop effects and we naturally expect $\lambda\sim N_{\rm TC}/(4\pi)^2$.
The S parameter was also calculated through the current correlators by the standard way.
We found  general tendency of the dependence of 
the meson masses relative to $F_\pi$, 
($M_\rho/F_\pi,\,M_{a_1}/F_\pi,\,M_{\rm TD}/F_\pi$) on $\gamma_m$, $S$, 
and $\Gamma$. 

  We find a characteristic feature of the techni-dilaton mass related to 
the conformality of SWC-TC:   
For fixed $S$ and $\gamma_m$,  absolute values of 
$(M_\rho/F_\pi)$ and $(M_{a_1}/F_\pi)$ are not sensitive to $\Gamma$, although they get {\it degenerate for large $\Gamma$}. On the contrary, 
{\it $(M_{\rm TD}/F_\pi)$ substantially decreases as $\Gamma$ increases}. Actually, in the formal limit $\Gamma \to \infty$
we would have   $(M_{\rm TD}/F_\pi)\to 0$ (This is contrast to the straightforward computation through ladder SD and BS equations mentioned before~\cite{Harada:2003dc}).   For fixed $S$ and $\Gamma$, 
again $(M_\rho/F_\pi)$ and $(M_{a_1}/F_\pi)$ are not sensitive to $\gamma_m$, while
$(M_{\rm TD}/F_\pi)$ substantially decreases as $\gamma_m$ increases.

 Particularly for the case of $\gamma_m = 1$, 
we study the dependence of the $S$ parameter on $(M_\rho/F_\pi)$ 
for typical values of $\Gamma$. 
It is shown that the techni-gluon contribution reduces the value of $S$ 
about 10\% in the region of $\hat S \lesssim 0.1$, 
although the general tendency is similar to 
the previous paper~\cite{Haba:2008nz} without techni-gluon condensation:  
$\hat{S}$ decreases to zero monotonically with respect to $(F_\pi/M_\rho)$. 
This implies $(M_\rho/F_\pi)$ necessarily increases when $\hat S$ is required
to be smaller.

To be more concrete, we consider a couple of typical models of SWC-TC 
with $\gamma_m \simeq1$ and $N_{\rm TC} = 2,3,4$ based on the CBZ-IRFP  
in the large $N_f$ QCD. 
Using 
the non-perturbative conformal anomaly Eq.(\ref{npanomaly}) together with the non-perturbative beta function Eq.(\ref{Miranskybeta}) and
Eq.(\ref{Miransky}), 
we find a concrete  relation between $\Gamma$ and  $(\Lambda_{\rm ETC}/F_\pi)$: 
In the case of $N_{\rm TC}=3$ ($N_f = 4 N_{\rm TC}$) and $S \simeq 0.1$,
we have $\Gamma \simeq 7$ for $(\Lambda_{\rm ETC}/F_\pi) = 10^4$--$10^5$
(required by the FCNC constraint).
Thanks to the large anomalous dimension $\gamma_m\simeq 1 $ and 
large techni-gluon condensation $\Gamma\simeq 7$, we obtain 
a relatively light techni-dilaton with mass 
\beq
M_{\rm TD} \simeq 600 \, \GeV
\label{prediction}
\eeq
compared with $M_\rho \simeq M_{a_1} \simeq 3.8 \, \TeV$ (almost degenerate).  Eq. (\ref{prediction}) is
consistent with the perturbative unitarity of $W_LW_L$ scattering even for large $M_\rho, M_{a_1}$. 
Note that largeness of $M_\rho$ and $M_{a_1}$ is essentially determined 
by the requirement of $S = 0.1$ fairly independently of techni-gluon 
condensation.~\footnote{The calculated  $S$ parameter here was from the TC dynamics alone and could be drastically 
changed by incorporating contributions from the generation mechanism of the mass of SM fermions such as the 
strong ETC dynamics. For instance, the fermion delocalization~\cite{Cacciapaglia:2004rb} in the Higgsless models as a possible analogue of certain ETC effects in fact  can cancel
large positive $S$ arising from  the 5-dimensional gauge sector which corresponds to the pure TC dynamics.  If it is the case in the explicit ETC model, then the overall mass scale of techni-hadrons including TD would be much lower than the  above estimate down to, say, 300 GeV. 
 }

The essential reason for the large $\Gamma$ 
is due to the existence of 
the wide conformal region $F_\pi (\sim m)  < \mu < \Lambda_{\rm ETC}$
with $\Lambda_{\rm ETC}/F_\pi = 10^4$--$10^5$,
which yields the smallness of the beta function 
(see Eq.({\ref{Miranskybeta}) and Eq. (\ref{Miransky}) ) and hence amplifies 
the techni-gluon condensation 
compared with the ordinary QCD with $\Gamma=1$.  Actually,  in the idealized (phenomenologically uninteresting) limit $\Lambda_{\rm ETC}/F_\pi \to \infty$,
we would have $\Gamma \to \infty$, which in turn would imply $M_{\rm TD}/F_\pi \to 0$ as mentioned above. \footnote{
This does not mean existence of true (exactly massless) NG boson of the broken scale symmetry, since  such a would-be NG boson gets decoupled in our case:
 The decay constant of techni-dilaton $F_{\rm TD}$ 
would diverge, $F_{\rm TD}/F_\pi  \to \infty$, in that limit through the PCDC relation $F_{\rm TD}^2 = -4 \langle  \theta^\mu_\mu\rangle/M_{\rm TD}^2
 \sim m^4/M_{\rm TD}^2 \sim F_\pi^4/M_{\rm TD}^2$. The scale symmetry is broken explicitly as well as spontaneously. 
}

To conclude,  various methods predicted the mass of the techni-dilaton (``conformal Higgs") in the range of $M_{\rm TD} \simeq 500-600$ GeV,
 which is within reach of LHC discovery. 

Since the SWC TC models are strong coupling theories and the ladder approximation/holographic calculations  would be no more than a qualitative hint,  more reliable calculations   are certainly needed, including the lattice simulations, before drawing a definite conclusion about the physics predictions.  Besides the phase diagram including the TC-induced/ETC-driven four-fermion couplings on the lattice, more reliable calculations such as the spectra as well as anomalous dimensions, non-perturbative
beta functions, 
$S$ parameter, etc.  are highly desired. 

\section*{Acknowledgments}
We thank K. Haba, M. Harada, M. Hashimoto, M. Kurachi, and S. Matsuzaki for collaborations on the relevant topics and useful discussions. 
 This work was supported by the JSPS Grant-in-Aid for Scientific Research(S) 
\#\ 22224003 and by the Nagoya University Foundation.

 \end{document}